\documentclass{article}
\usepackage[utf8]{inputenc}
\usepackage{authblk}
\usepackage{setspace}
\usepackage[margin=1.25in]{geometry}
\usepackage{graphicx}
\usepackage{xr}
\graphicspath{./figs/}
\usepackage{subcaption}
\usepackage{amsmath}
\usepackage{parskip}

\usepackage{amssymb}
\usepackage{lineno}
\usepackage{multirow}
\usepackage{tabularx}
\usepackage{siunitx}
\usepackage{float}
\usepackage{booktabs}
\usepackage{multirow}

\usepackage[style=ieee, sorting=none, citestyle=numeric-comp]{biblatex} 
\bibliography{./ref2}

\title{Beyond defect thermodynamics: first-principles carrier capture at interstitial alkaline-earth dopants in kesterite Cu$_2$ZnSnS$_4$}

\author[1]{Anarga Paritrana Widhi}
\author[1,2*]{Mohammad Kemal Agusta}
\author[3]{Kenta Hongo}
\author[2,4]{Abdul Rohman Supandi}
\author[1,2**]{Ganes Shukri}

\affil[1]{Nano and Quantum Technology Research Group, Faculty of Industrial Technology, Bandung Institute of Technology, Bandung 40132, West Java, Indonesia.}
\affil[2]{Research Centre for Nanoscience and Nanotechnology (RCNN), Bandung Institute of Technology, Bandung 40132, West Java, Indonesia.}
\affil[3]{Center for Advanced Scientific Computing, Japan Advanced Institute of Science and Technology (JAIST), Nomi, Ishikawa-ken, Japan}
\affil[4]{Division of Inorganic and Physical Chemistry, Faculty of Mathematics and Natural Sciences, Bandung Institute of Technology, Bandung 40132, West Java, Indonesia.}
\affil[*]{Corresponding author: kemal@itb.ac.id}
\affil[**  ]{Corresponding author: ganes$\textunderscore$tf07@itb.ac.id}

\date{}

\begin{document}

\maketitle

\begin{abstract}
The kesterite Cu$_{2}$ZnSnS$_{4}$ (CZTS) is a promising earth-abundant photovoltaic absorber due the abundance of its constituent elements and low toxicity. However, its performance remains limited by severe open-circuit voltage losses mainly due to defect-assisted nonradiative recombination. Although alkaline-earth alloying has been proposed as an effective strategy for suppressing Cu–Zn antisite disorder and mitigating band-tailing effects, the impact of alkaline-earth-related defects on carrier recombination remains largely unexplored. Here, we employ first-principles nonradiative multiphonon emission theory to investigate the carrier-capture activity of Be, Mg, and Ca dopants in CZTS. Electronic structure calculations reveal the formation of defect-induced states due to interstitials in which the energetic positions and localization characteristics depend strongly on the alloying species. Configuration-coordinate analyses show substantial differences in lattice relaxation associated with charge-state transitions, leading to noticable different carrier-capture behaviors. Be and Mg interstitials exhibit significant nonradiative carrier-capture activity, while Ca interstitials possess substantially smaller carrier-capture coefficients and are therefore less detrimental to carrier lifetime. In particular, Mg interstitials provide efficient electron and hole capture pathways capable of supporting complete Shockley–Read–Hall (SRH) recombination cycles. The contrasting recombination behavior is attributed to the differences in defect electronic structure and defect-induced lattice relaxation. These results demonstrate that suppressing Cu–Zn disorder alone is insufficient for assessing alloying strategies in CZTS and highlight the importance of directly evaluating carrier-capture processes. In a broader context, our work establishes carrier-capture coefficients as a critical descriptor for defect engineering and identifies Ca as a comparatively benign alloying element for minimizing nonradiative recombination losses in kesterite CZTS photovoltaic absorbers.
\end{abstract}

\section{Introduction}
The kesterite Cu$_2$ZnSnS$_4$ (CZTS) has attracted considerable attention as a thin-film photovoltaic absorber owing to its suitable direct band gap, high optical absorption coefficient, and composition of earth-abundant, environmentally benign elements \cite{tangjiao}. These properties make it a promising alternative to established thin-film technologies such as Cu(In,Ga)Se$_2$ and CdTe \cite{tangjiao, Hadke2025}. Despite a decade of intensive development, however, the power conversion efficiency of CZTS solar cells remains well below its theoretical limit, primarily because of a severe open-circuit voltage ($V_{oc}$) deficit \cite{Hadke2025, efflimit, li2022unveiling}.

The origin of this $V_{oc}$ deficit has been studied extensively. Early work identified Cu--Zn disorder and the associated Cu$_\mathrm{Zn}$ and Zn$_\mathrm{Cu}$ antisite defects as major sources of band tailing, electrostatic potential fluctuations, and recombination losses \cite{gokmen2013, werner2005}. Because Cu$^{+}$ and Zn$^{2+}$ have similar ionic radii, these antisite pairs form with unusually low energy, leading to substantial cation disorder in the kesterite lattice \cite{chen2012APL, bourdais2016cu}. Considerable effort has therefore been devoted to suppressing Cu--Zn disorder through compositional engineering and cation substitution \cite{ujiayu2025, sharif2020aem, chen2013advmat}.

Partial substitution of Zn by alkaline-earth elements has been proposed as a way to reduce Cu--Zn disorder in CZTS. In our previous study, the formation energy of the Cu$_\mathrm{Zn}$+Zn$_\mathrm{Cu}$ antisite pair increased at selected Be, Mg, and Ca substitution levels, with the trend depending on composition \cite{ujiayu2025}. That study also identified in-gap states for neutral alkaline-earth interstitials and found interstitial incorporation generally less favorable than substitution. Experiments on Mg incorporation have reported improved carrier transport at low Mg content and degraded device performance at higher Mg content \cite{steneracsami19}. These observations motivate examining whether defects introduced during alloying can provide additional carrier-capture pathways.

Previous first-principles studies of native defects in CZTS have shown that lattice relaxation and electron--phonon coupling can make particular Sn-related defects as efficient nonradiative capture centers \cite{killerdefect,lonepair}. Modeling their effect on device performance requires combining the capture coefficients with defect populations and carrier densities \cite{efflimit,Kim2020upperlimit}. For alkaline-earth interstitials, the presence of a gap state identifies a candidate for carrier capture, while the rates of the individual electron- and hole-capture steps determine whether a recombination pathway is plausible. 

Here, we extend our earlier thermodynamic and electronic-structure study \cite{ujiayu2025} by calculating charge-state-resolved carrier capture at Be, Mg, and Ca interstitials in CZTS. We use the nonradiative multiphonon formalism \cite{alkauskasprb,turianskycpc} to examine how the energy surfaces, lattice relaxation, and electronic coupling influence electron and hole capture. The comparison illustrates how capture kinetics can complement defect energetics when screening alloying elements.
%
\section{Computational Details}

\subsection{Electronic Structure and Defect Energetics}
\label{sec:methods_dft}

Spin-polarized density functional theory (DFT) calculations were performed using the projector augmented-wave (PAW) method as implemented in the Vienna Ab-Initio Simulation Package (VASP) \cite{hafner2008vasp, kresse1996, kresse1999}. The following PAW valence configurations were used: Cu($3d^{10}4p^{1}$), Zn($3d^{10}4p^{2}$), Sn($5s^{2} 4d^{10} 5p^{2}$), S($3s^{2}3p^{4}$), Be($2s^{2}$), Mg($3s^{2}$), and Ca($3s^{2}3p^{6}4s^{2}$). Exchange and correlation were treated within the generalized gradient approximation of Perdew, Burke, and Ernzerhof (GGA-PBE) \cite{PBE}, and the Kohn--Sham wave functions were expanded in a plane-wave basis with a kinetic energy cutoff of 520~eV. To correct the well-known band-gap underestimation of semilocal functionals in kesterite CZTS, a Hubbard-$U$ correction in the rotationally invariant formulation of Dudarev \textit{et al.} \cite{dudarev1998} was applied, with $U = 5.2$, $6.5$, and $3.5$~eV on the Cu-$3d$, Zn-$3d$, and Sn-$4d$ orbitals, respectively. The Hubbard-$U$ parameters were adopted from our previous study of Cu--Zn disorder in CZTS \cite{ujiayu2025}. The calculated host band gap and its implications for carrier capture are discussed in Sec.~\ref{sec:elec}.

The kesterite CZTS host was modeled with a $2\times2\times1$ supercell containing 64 atoms, with optimized lattice parameters $a = b = 10.879$~\AA{} and $c = 10.870$~\AA{} (Fig. S1). Brillouin-zone integrations were performed with a $\Gamma$-centered $4\times4\times4$ $k$-point mesh. Structural relaxations used the conjugate-gradient algorithm with a Gaussian smearing of 0.03~eV, and were considered converged when the residual force on every atom fell below 0.02~eV/\AA{}; the self-consistent electronic loop was converged to $10^{-5}$~eV. Electronic structures were subsequently evaluated at the relaxed geometries using the tetrahedron method with Bl\"ochl corrections. Band structures and projected densities of states were generated with Pymatgen \cite{pymatgen}, and structures were visualized with VESTA \cite{VESTA}.

Dopant atoms were introduced at the most favorable sites identified in our previous work\cite{ujiayu2025}. Interstitial dopant sites are located between the Sn and Zn planes and coordinated by four neighbouring S atoms. Cu-substitutional dopants are located on the Sn plane. Different charge states were considered for the different sites For the interstitial site, the neutral ($q=0$), singly charged ($q=+1$), and doubly charged ($q=+2$) states were considered, while for the Cu-substitutional site,  the neutral and singly charged states were considered. For the Zn-substitution, only the $q=-1,0,+1$ charge states are evaluated. The charge state was set through the total electron count, with a compensating uniform background for charged cells. Each charge state was relaxed independently while keeping the lattice vectors fixed at their optimized bulk values similar to our previous study \cite{prima2023effect}. The carrier-capture calculations were then performed on the interstitial configurations.

The formation energy of an interstitial defect in charge state $q$ was evaluated as
\begin{equation}
		E_\mathrm{form}(X^{q}) = E_\mathrm{tot}(X^{q}) - E_\mathrm{tot}(\mathrm{bulk}) - \sum_i n_i \mu_i + q\, 	
		(E_F - E_\mathrm{VBM}) + E_\mathrm{corr},
	\label{eq:eform}
\end{equation}
where $E_\mathrm{tot}(X^{q})$ and $E_\mathrm{tot}(\mathrm{bulk})$ are the total energies of the defective and pristine supercells, $n_i$ and $\mu_i$ are the number and chemical potential of the added species, $E_F$ is the Fermi level referenced to the valence-band maximum $E_\mathrm{VBM}$, and $E_\mathrm{corr}$ is the electrostatic finite-size correction for charged supercells \cite{freysoldt2014}. The latter was treated at the level of the first-order Makov--Payne (point-charge) scheme \cite{MakovPayne1995},
	\begin{equation}
		E_\mathrm{corr} = \frac{q^{2}\alpha}{2\varepsilon L},
	\label{eq:makov}
	\end{equation}
with $\alpha$ the Madelung constant, $L$ the supercell length, and $\varepsilon$ the static dielectric constant, assumed isotropic. The thermodynamic charge-transition level $\varepsilon(q_1/q_2)$ was then obtained as the Fermi energy at which the formation energies of the two charge states are equal,
	\begin{equation}
		\varepsilon(q_1/q_2) =
		\frac{E_\mathrm{form}(X^{q_1};\,E_F=0) - E_\mathrm{form}(X^{q_2};\,E_F=0)}
		{q_2 - q_1}.
	\label{eq:tll}
	\end{equation}

\subsection{Nonradiative Carrier-Capture Calculations}
\label{sec:nonrad_method}
We modeled nonradiative carrier capture at Be$_i$, Mg$_i$, and Ca$_i$ using the one-dimensional configuration-coordinate (CC) approximation to multiphonon capture \cite{HuangRhys1950,henrylang1977,passler1976,alkauskasprb,turianskycpc}. The generalized configuration coordinate Q describing the collective atomic displacement between the equilibrium geometries of two defect charge states is defined as

	\begin{equation}
		Q=\left[\sum_{\alpha}m_{\alpha}\left(\mathbf{R}^{(i)}_{\alpha}-\mathbf{R}^{(f)}_{\alpha}\right)^2\right]^{1/2},
		\label{eq:Q}
	\end{equation}

where \(m_{\alpha}\) is the mass of atom \( \alpha \), while \( \mathbf{R}^{(i)}_{\alpha} \) and \( \mathbf{R}^{(f)}_{\alpha} \) denote the atomic coordinates in the initial and final charge states, respectively. 

We sampled the potential-energy surfaces using single-point calculations at geometries obtained by linear interpolation and extrapolation between relaxed endpoints. Harmonic fits were used for the approximately parabolic surfaces. For the anharmonic surfaces, we used spline potentials and solved the one-dimensional vibrational problem numerically with CarrierCapture.jl \cite{KimAnharmonic2019,kim2020carriercapture}. Where a classical intersection was identified along the sampled path, the capture barrier was measured relative to the minimum of the initial surface.

The electron-phonon coupling matrix element associated with carrier capture is given by

	\begin{equation}
		W_{if} = \left<\psi_i\left | \frac{\partial \hat h}{\partial Q} \right|\psi_f\right>,
		\label{eq:Wif}
	\end{equation}

where \( \psi_i \) and \( \psi_f \) are the electronic wave functions of the initial and final states, respectively, and \( \hat{h} \) is the Kohn-Sham Hamiltonian for independent-particle picture. The quantity \( W_{if} \) provides a measure of the coupling strength between the electronic transition and the lattice distortion accompanying the charge-state transition. In practice, Eq. ~(\ref{eq:Wif}) was evaluated using first-order perturbation theory as $W_{if} = (\epsilon_f - \epsilon_i)\,\langle \psi_i | \partial \psi_f / \partial Q \rangle$, where $\epsilon_i$ and $\epsilon_f$ are the Kohn--Sham eigenvalues of the initial and final states. The wave-function overlaps were obtained from the WSWQ routine of VASP at the $\Gamma$ point, using a set of structures displaced by small amounts about points with a midgap state.

The magnitude of lattice relaxation associated with the carrier-capture process was further characterized by the Huang-Rhys factor \(S\),

	\begin{equation}
		S=\frac{\Omega (\Delta Q)^2}{2\hbar},
	\end{equation}

where \( \Omega \) is the effective phonon frequency. Large Huang-Rhys factors indicate strong electron-phonon coupling and substantial lattice relaxation, both of which are known to facilitate efficient nonradiative multiphonon transitions.

The carrier-capture coefficient \( C_{n,p} \) was then evaluated within the first-principles NMP formalism according to

	\begin{equation}
		C_{n,p} = Vr_{n,p},
		\label{eq:C}
	\end{equation}

where \(V\) is the supercell volume and \(r_{n,p}\) is the carrier-capture rate obtained from Fermi’s Golden Rule. The latter depends on the electron-phonon coupling matrix elements, vibrational overlap integrals, and thermal occupation of phonon states. Detailed expressions for the carrier-capture rate can be found in the following reference \cite{alkauskasprb, turianskycpc}. The capture coefficients for charged defects were scaled by the temperature-dependent Sommerfeld factor $s(T)$, which enhances or suppresses capture at attractive or repulsive centers, respectively. The Sommerfeld factor was evaluated within the effective-mass approximation using the static dielectric constant of CZTS taken from experiment \cite{sharmin2020}, together with the electron and hole effective masses obtained from previous first-principles calculations of pristine CZTS \cite{liu2012}; the dielectric constant and effective masses of the defective supercells were assumed identical to those of the pristine host. A charge state scaling correction accounting for the finite supercell size was additionally applied, following the implementation in \texttt{nonrad} \cite{turianskycpc}. The electron and hole capture coefficients were then computed for all relevant charge state transitions over the temperature range 25--800~K.

The calculated configuration-coordinate diagrams, electron-phonon coupling strengths, energy barriers, and carrier-capture coefficients were subsequently used to assess the nonradiative recombination activity of the investigated defects and their potential role as Shockley-Read-Hall recombination centers in CZTS.

\section{Result \& Discussion}

\subsection{Electronic Structure \& Transition Energy Levels}
\label{sec:elec}
The electronic structure of a point defect provides the first, but only a partial, indication of its influence on carrier recombination. Within the Shockley--Read--Hall (SRH) picture, a defect can act as a recombination centre only if it introduces a localized state inside the band gap that is energetically accessible to both the conduction and valence bands. The existence of such a state is therefore a \emph{necessary} condition for nonradiative recombination; whether the defect is an \emph{efficient} recombination centre is decided by the carrier-capture probability, which depends on the lattice relaxation and electron--phonon coupling analysed in sections ~\ref{sec:cc} and \ref{sec:capture}. Here, we first establish the electronic structure and thermodynamic charge-transition levels of the alkaline-earth dopants to identify the sites, dopants, and charge states which enable non-radiative recombination. 

\begin{figure}[h]
    \centering
    \includegraphics[width=0.75\textwidth]{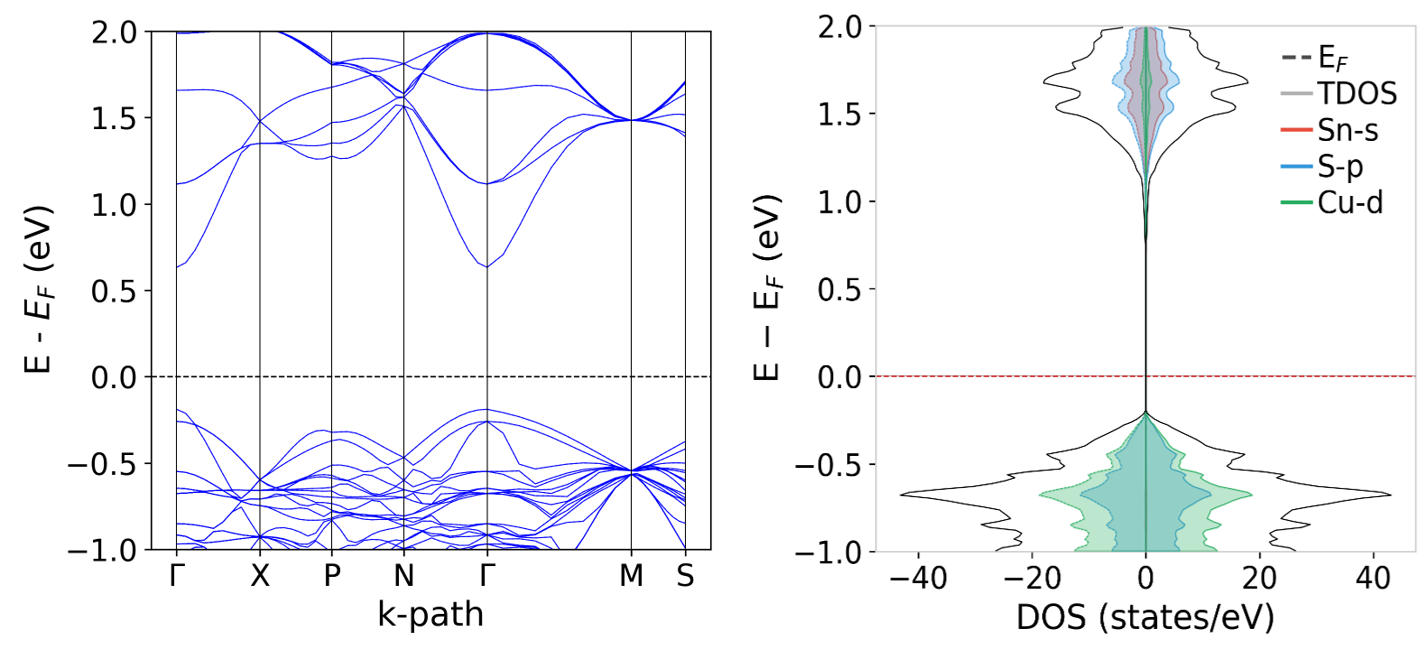}
    \caption{Electronic band-structure and projected density of states (PDOS) of pristine kesterite Cu$_2$ZnSn$S_4$.}
    \label{fig:1}
\end{figure}

Figure~\ref{fig:1} shows the band structure and projected density of states (PDOS) of pristine kesterite CZTS. Consistent with previous first-principles studies, the valence-band maximum (VBM) is dominated by hybridized Cu-$3d$ and S-$3p$ states, whereas the conduction-band minimum (CBM) is derived mainly from Sn-$5s$ states with a smaller S-$3p$ contribution \cite{ujiayu2025, berman2019role}. The calculated PBE$+U$ band gap is 0.82~eV, which underestimates the experimental value of $\sim$1.5~eV, as expected for a semilocal functional even with an on-site Hubbard correction. Because the carrier-capture coefficient depends on the position of the defect level relative to the band edges, this underestimation introduces a quantitative uncertainty in the absolute capture rates. We therefore use the calculated coefficients as model-dependent estimates to compare the carrier-capture pathways. 


Despite a neutral alkaline-earth metal not being isovalent with Cu, no deep defect state is introduced by all dopants at the Cu-substitutional site in both the neutral and singly charged states. The thermodynamic transition levels indicate that the neutral dopants provide a shallow defect state resonant with the conduction band. Thermodynamically, the +1 charge state is shown to be stable for the majority of the band gap, with the transition occuring above the conduction band minimum for all dopants (fig. S14). In all cases, the dopant atom is not significantly displaced between the different charge states. Similiarly, Zn-substitutional dopants also do not result in deep defect states, which is expected due to the isovalence. The neutral charge state is found to be stable throughout the band gap for all dopant elements (fig. S15). Thus, within the SRH picture, all dopants on both sites will be unable to act as non-radiative recombination centers.

\begin{figure}[h]
    \centering
    \includegraphics[width=0.75\textwidth]{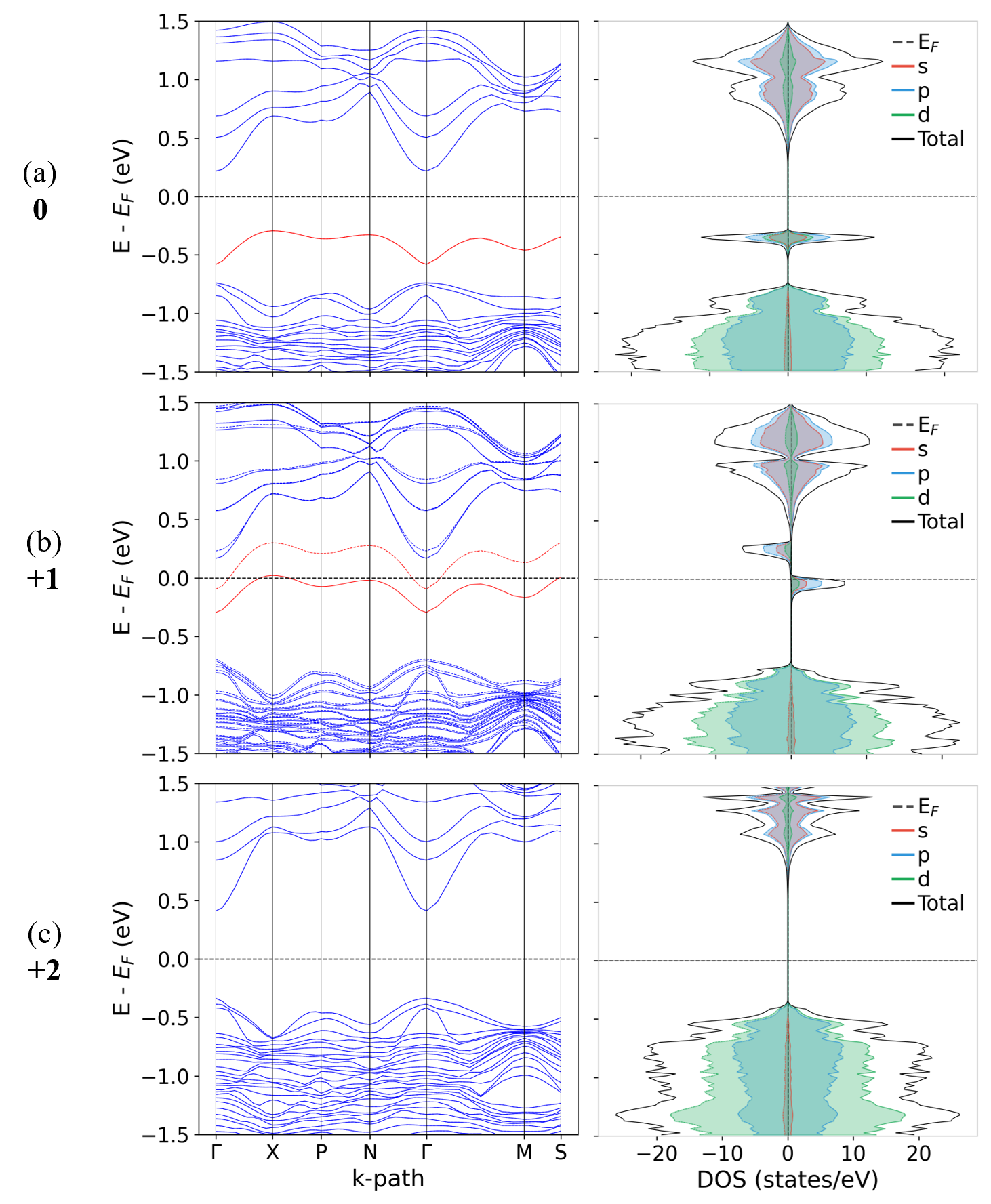}
    \caption{Electronic band-structure and projected density of states (PDOS) of Mg$_i$-doped kesterite Cu$_2$ZnSn$S_4$ in: (a) neutral, (b) +1 and (c) +2 charge state, respectively.}
    \label{fig:2}
\end{figure}

 Introducing an interstitial Be, Mg, or Ca atom produces a localized state deep in the gap in the neutral charge state (Fig.~\ref{fig:2}a, S5a and S6a). The defect bands of all three dopants are flat throughout the Brillouin zone, implying a localized, point-defect character. Their evolution with charge state is likewise common to all three dopants. In the neutral state the gap level is fully occupied; upon removal of one electron (the singly charged state) it becomes spin-split, producing an exchange splitting of $\approx$0.28~eV for Be$_i$ and $\approx$0.20~eV for Mg$_i$ and Ca$_i$, with the unoccupied spin-down level pushed toward the CBM (Fig.~\ref{fig:2}b, S5b and S6b). Whereas, in the equilibrium position of the doubly charged state no defect state is present in the band gap, restoring a nonmagnetic configuration (Fig.~\ref{fig:2}c, S5c and S6c). However, site-projected density of state analysis shows that for the Ca and Mg interstitial-doped cases a somewhat localized defect state is present slightly above the CBM, which is not observed in the Be-interstitial case (fig. S11-S13).

\begin{figure}[h]
    \centering
    \includegraphics[width=0.8\textwidth]{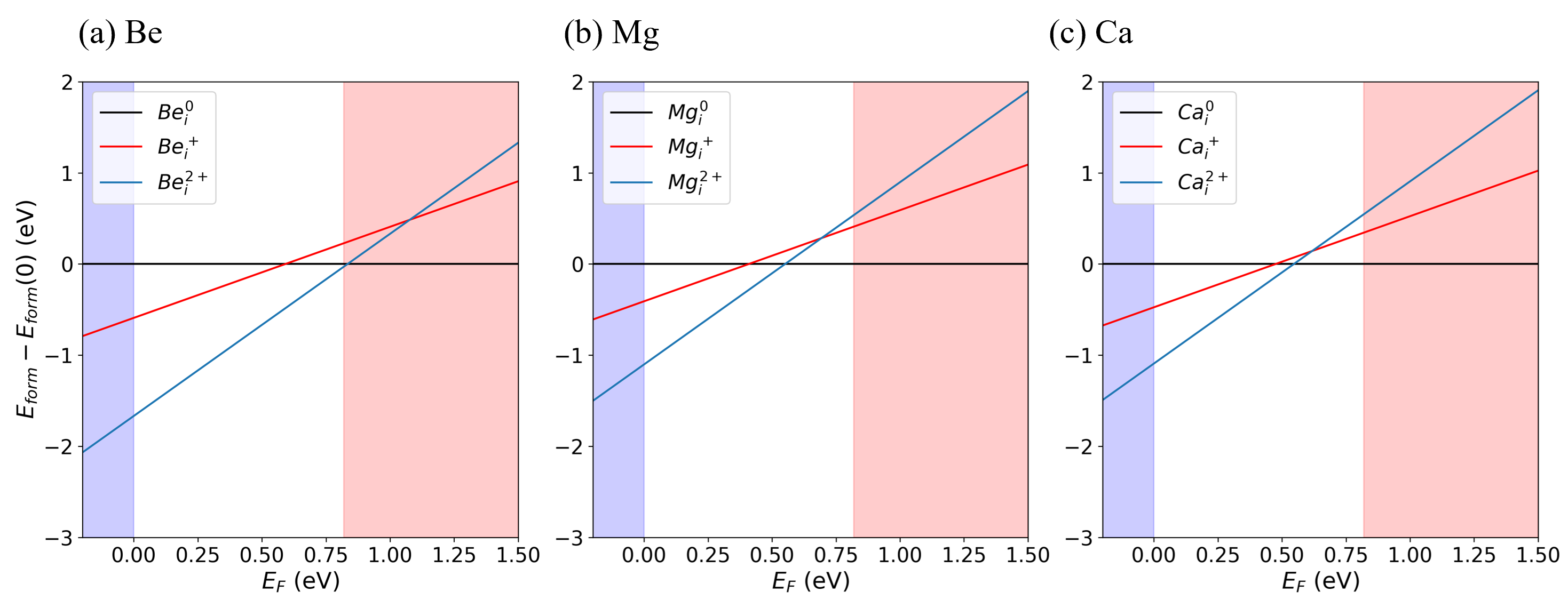}
    \caption{Calculated defect formation energies of (a) Be$_i$, (b) Mg$_i$, and (c) Ca$_i$ interstitial defects in CZTS as a function of the Fermi level under thermodynamic equilibrium. The black, red, and blue lines represent the neutral, singly charged, and doubly charged charge states, respectively. The slope of each line corresponds to the defect charge state, while the intersections between adjacent charge-state lines define the thermodynamic charge transition levels. The shaded blue and red regions denote the valence-band maximum (VBM) and conduction-band minimum (CBM), respectively.}
    \label{fig:3}
\end{figure}

The thermodynamic charge-transition levels, obtained from the defect formation energies as a function of the Fermi level, are shown in Fig.~\ref{fig:3}.  For Be$_i$, the doubly charged state is the most stable configuration almost across the entire gap: both valence electrons of the interstitial are donated to the host, and no neutral or singly charged region appears. For Mg$_i$ and Ca$_i$, the neutral and doubly charged states are the stable configurations at low and high Fermi energy, respectively, with a direct $(+2/0)$ transition located at 0.55~eV above the VBM in both cases. The singly charged state is metastable for all three dopants: the $(+1/0)$ level lies above the $(+2/+1)$ level (a negative-$U$ ordering), so $+1$ is never the thermodynamic ground state at any Fermi energy. A metastable charge state nonetheless remains kinetically relevant, because it is transiently populated during carrier cycling and mediates the individual single-carrier capture events of an SRH cycle \cite{alkauskasprb}. Therefore, we also evaluate the two single-electron transitions $(+1/0)$ and $(+2/+1)$, which together constitute the possible sequential recombination pathways, rather than the direct two-electron $(+2/0)$ transition, which would require the improbable simultaneous capture of two carriers.

Taken together, the electronic-structure analyses establish the necessary conditions for the SRH recombination: all three alkaline-earth interstitials introduce deep, host-derived gap states with comparable charge-transition levels. In contrast, the dopants in the Cu-substitutional and Zn-substitutional sites do not possess the conditions for SRH recombination.

%

%

\subsection{Configuration Coordinate Diagram}
\label{sec:cc}
\begin{figure}[h]
    \centering
    \includegraphics[width=0.9\textwidth]{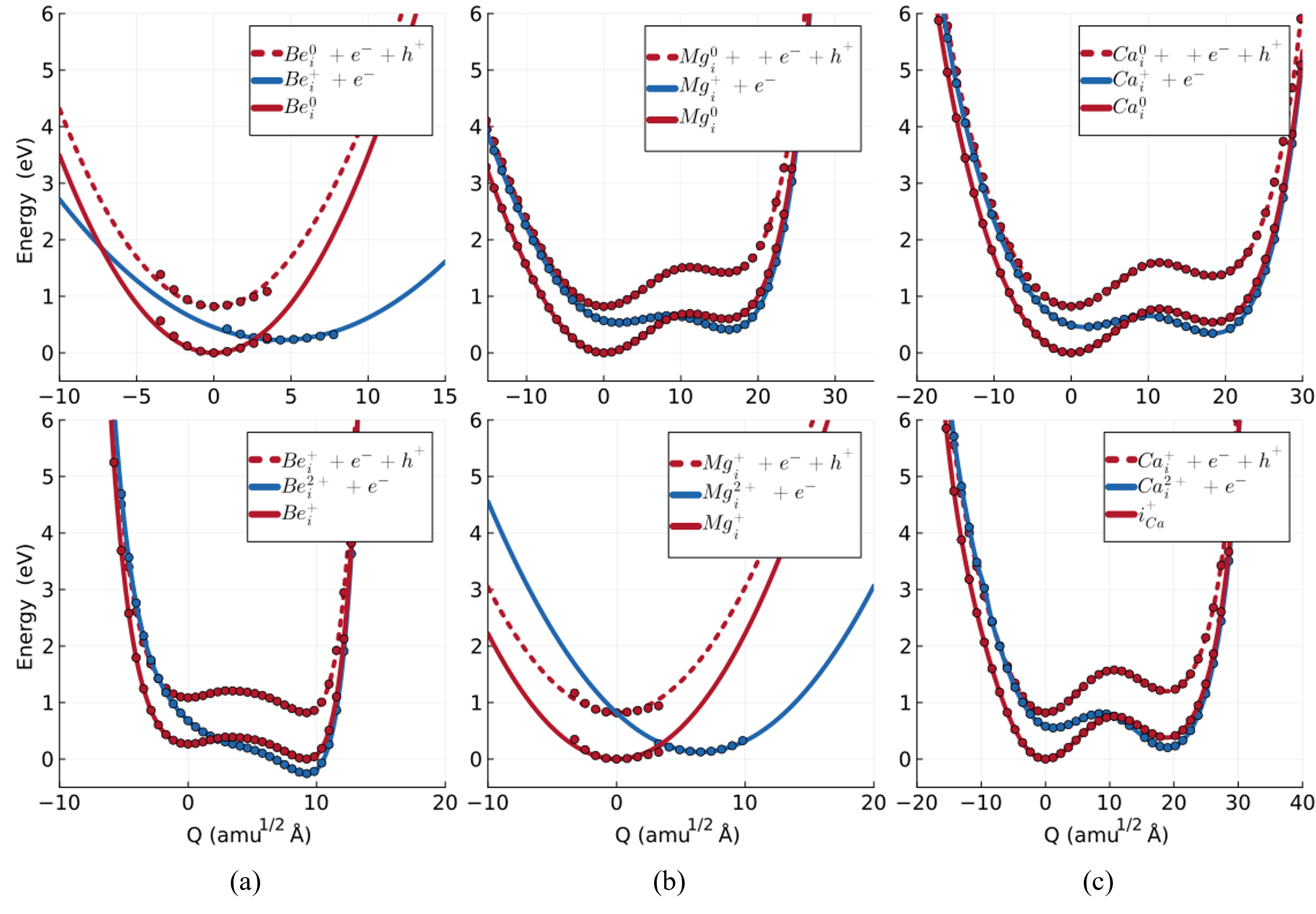}
    \caption{Configuration-coordinate (CC) diagrams for the carrier-capture transitions associated with (a,b) Be$_i$, (c,d) Mg$_i$, and (e,f) Ca$_i$ interstitial defects in CZTS. The dots represent DFT-calculated total energies along the generalized configuration coordinate \(Q\), while the  curves are harmonic and anharmonic fits to the potential energy surfaces of the corresponding defect charge states. The horizontal displacement between the minima reflects the lattice relaxation accompanying charge-state transitions, whereas the vertical separation corresponds to the defect excitation energy.}
    \label{fig:4}
\end{figure}

The electronic-structure analysis of sec.~\ref{sec:elec} establishes that all three interstitials introduce deep gap states with comparable charge-transition levels, satisfying the necessary electronic condition for SRH recombination. Whether a given charge-state transition supports efficient nonradiative capture, however, is determined by the lattice relaxation that accompanies the electronic transition, which we describe within the configuration-coordinate (CC) formalism, according to the method described in sec.~\ref{sec:nonrad_method}. The classical energy barriers ($E_b$) for nonradiative carrier capture are obtained from this calculation. The resulting diagrams for the $(+1/0)$ and $(+2/+1)$ transitions of Be$_i$, Mg$_i$, and Ca$_i$ are shown in Fig.~\ref{fig:4}, and the corresponding $\Delta Q$ values are collected in Table~\ref{tab:dQ}. The anharmonic approach was necessary for four out of six transitions.

The lattice response differs among the three dopants and between the two transitions of each dopant. The configurational displacement spans a factor of roughly five, from $\Delta Q = 4.3$~amu$^{1/2}$\,\AA{} for the Be$_i$ $(+1/0)$ transition to $\Delta Q = 19.0$~amu$^{1/2}$\,\AA{} for the Ca$_i$ $(+2/+1)$ transition. This ordering generally follows the local structural perturbation imposed by each interstitial: the small Be atom produces a compact distortion confined to its immediate sulfur coordination shell, whereas the larger Mg and Ca atoms displace a broader region of the Sn--S and Cu--S bonding network, yielding a larger $\Delta Q$, with the Ca-doped system.

The CC diagrams reveal qualitatively different topologies for each transition. For Be$_i$, the small $(+1/0)$ displacement places the crossing point of the electron-capture branch essentially at the minimum of the initial surface, so that electron capture is nearly barrierless. The hole-capture branch of the same transition, however, exhibits no intersection between the two potential-energy surfaces, indicating that no thermally activated multiphonon pathway exists for hole capture; the corresponding $0\!\rightarrow\!+1$ process can proceed only radiatively, through an optical transition that emits a photon. Consequently, the Be$_i$ $(+1/0)$ transition can trap an electron efficiently but cannot complete a nonradiative recombination cycle. The Be$_i$ $(+2/+1)$ transition, by contrast, involves a much larger displacement and presents an anharmonic vibrational characteristic with relatively high capture barriers.

For Mg$_i$, the situation is reversed between the two transitions: the $(+1/0)$ transition shows a large displacement and anharmonicity but without a hole-capture crossing, whereas the $(+2/+1)$ transition has a moderate displacement, comparatively flat crossing geometry, and low barriers for both electron and hole capture, making it the channel most favourable to nonradiative recombination in the classical picture. For Ca$_i$, both transitions are characterized by the largest displacements of the entire set and by particularly soft, anharmonic potential energy surfaces. Similiar with other $(+1/0)$ transitions, there is no hole-capture crossing.  

All transitions show  Huang-Rhys factor values $S\gg1$, indicating nonradiative capture centers with large lattice relaxations. For example, the Mg(+2/+1) transition has a Huang-Rhys factor value of 59 in its initial state, larger than the values for several defects with harmonic recombination characteristics such as $\mathrm{GaN:C_N}$ and $\mathrm{ZnO:Li_{Zn}}$ \cite{alkauskasprb}. However large lattice relaxation does not necessarily imply higher carrier capture rate, as it could lead to relatively high capture barriers, as shown in table \ref{tab:dQ}. Furthermore, in this approach, the capture rate also depends on the electronic wavefunction overlap (eq. \ref{eq:Wif}).

\begin{table*}[t]
\centering
\caption{Configurational displacements $\Delta Q$ and classical energy barriers $E_b$ of alkaline-earth interstitial defects in CZTS.}
\label{tab:dQ}
\begin{tabular}{llccc}
\toprule
Defect & Charge-state transition &  $\Delta Q$ (amu$^{1/2}$\,\AA)&
$E_b(\mathrm{e})$ (eV) &
$E_b(\mathrm{h})$ (eV) \\
\midrule
\multirow{2}{*}{Be$_i$}
& $+1/0$ & $4.32$ &$0.031$ & --- \\
& $+2/+1$ & $9.23$ &$0.621$ & $0.596$ \\
\midrule
\multirow{2}{*}{Mg$_i$}
& $+1/0$ & $16.2$ &$0.244$ & --- \\
& $+2/+1$ & $6.54$&$0.148$ & $0.000$ \\
\midrule
\multirow{2}{*}{Ca$_i$}
& $+1/0$ &$18.3$ &$0.29$ & --- \\
& $+2/+1$ &$19.0$ &$0.556$ & $1.495$ \\
\bottomrule
\end{tabular}
\end{table*}
%
%

%
\subsection{Carrier Capture Coefficients}
\label{sec:capture}
In this section, we discuss the calculated temperature-dependent capture coefficients $C_{n,p}$ within the NMP formalism (Eq. ~(\ref{eq:C}) for the $(+1/0)$ and $(+2/+1)$ transitions of each dopant. The results at $T=300$~K are summarized in Table ~\ref{tab:capture}. The full temperature dependence over the range 25--800~K along with the calculated Sommerfeld factors are shown in Fig. S22-S23 of the Supporting Information. 

For the $(+1/0)$ transition, the three dopants differ significantly in their electron-capture behavior. Be$_i$ exhibits the largest electron-capture coefficient, $C_n \approx 2.38\times10^{-9}$~cm$^3$\,s$^{-1}$ at 300~K, several orders of magnitude than the Mg$_i$ and Ca$_i$ (viz. $2.03\times10^{-12}$ and $1.78\times10^{-13}$~cm$^3$\,s$^{-1}$, respectively). The large calculated capture-coefficient follows from the near-barrierless electron-capture geometry of Be$_i$ (Sec.~\ref{sec:cc}) with its strong conduction-band coupling. This finding also indicates that neutral Be$_i$ is potentially an efficient electron trap under operating conditions. It is however still two orders of magnitude below the electron-capture coefficient reported for the Sn$_\mathrm{Zn}$, one of the recognized recombination centres in CZTS \cite{lonepair}. Although much slower, the Mg$_i$ and Ca$_i$ can still be considered as active electron capture centers, as their rates are still comparable to the typical radiative recombination coefficients. As established in Sec.~\ref{sec:cc}, all $(+1/+0)$ transition has no nonradiative hole capture pathway, so these electron-capture rates cannot be converted into a complete recombination cycle by multiphonon emission alone.

The ordering changes completely for the $(+2/+1)$ transition. Here Mg$_i$ becomes dominant, with electron- and hole capture coefficients of $C_n \approx 7.14\times10^{-10}$ and $C_p \approx 1.25\times10^{-11}$~cm$^3$\,s$^{-1}$, both of which well above typical semiconductor radiative capture values of $\sim\!10^{-13}$~cm$^3$\,s$^{-1}$\cite{alkauskasprb}, thus making non-radiative capture dominant. These values are however still two to three orders of magnitude lower than the capture cross sections reported for the Sn$_\mathrm{Zn}$ and Cu$_\mathrm{Sn}$ antisites \cite{lonepair}, indicating a moderately active recombination center. In contrast,  both the Be$_i$ and Ca$_i$ coefficients collapse for this transition to values orders of magnitude below the radiative values, reflecting their higher barriers compared to their $(+1/0)$ transitions. 

These results demonstrate that the dominant capture pathway is a property of the individual charge-state transition rather than of the chemical identity of the dopant, and that recombination activity cannot be inferred from a single large capture coefficient. Within the SRH mechanism, nonradiative recombination requires the defect to capture both an electron and a hole in sequence through its metastable $+1$ state; a defect that captures only one carrier type efficiently acts as a trap rather than a recombination centre. Be$_i$ captures electrons efficiently through its $(+1/0)$ transition but provides no nonradiative hole-capture pathway and is inactive for the subsequent $(+2/+1)$ transition, and therefore behaves as an electron trap rather than a full recombination centre. Ca$_i$ exhibits relatively small coefficients for both transitions and is the least active of the three dopants. Only Mg$_i$ maintains appreciable coefficients for both electron and hole capture (through its $(+2/+1)$ transition). Thus it provides a complete Shockley--Read--Hall recombination pathway, making it the most detrimental of the alkaline-earth interstitials among the three considered dopants.
%
%
\subsection{Discussion}
\label{sec:discussion}
The results above establish carrier capture as the critical factor for alkaline earth interstitials in CZTS. All three dopants introduce deep, host-derived gap states with closely spaced charge-transition levels (section ~\ref{sec:elec}), yet their capture coefficients range from efficient ($\gtrsim 10^{-9}$~cm$^3$\,s$^{-1}$) to negligible (far below the $\sim 10^{-13}-10^{-14}$~cm$^3$\,s$^{-1}$ coefficients of radiative transitions), a difference of many orders of magnitude (section ~\ref{sec:capture}). 

This distinction has direct consequences for the interpretation of alkaline-earth alloying. Our previous work showed that Be, Mg, and Ca alloying raises the formation energy of the Cu$_\mathrm{Zn}$ and Zn$_\mathrm{Cu}$ antisite pairs, suppressing the cation disorder responsible for band tailing and the open-circuit-voltage deficit \cite{ujiayu2025}. Experimentally, low-level Mg incorporation improves crystallinity, grain size, and carrier transport and enhances device efficiency, whereas excessive Mg degrades performance through reduced carrier lifetime and secondary-phase formation \cite{steneracsami19}. The present calculations offer a microscopic explanation for this concentration-dependent behavior based on the distinct roles of the two Mg lattice sites. Substitutional Mg on the Zn site suppresses Cu--Zn antisite formation and is beneficial, and dominates at low Mg content. Under Mg-rich growth conditions, however, the population of interstitial Mg increases, and Mg$_i$ provides a complete Shockley--Read--Hall pathway through its $(+2/+1)$ transition, with moderately high electron and hole capture coefficients. The experimentally observed optimum Mg concentration thus emerges naturally as a balance between the disorder suppression provided by substitutional Mg and the nonradiative recombination introduced by interstitial Mg.

The remaining two dopants illustrate the same principle from opposite directions. Be$_i$ captures electrons efficiently but lacks a nonradiative hole-capture channel, so it behaves as an electron trap rather than a recombination centre; its practical relevance is further limited by the toxicity of beryllium, and it is retained here primarily to establish the mechanistic link between electronic structure and capture activity. Ca$_i$ shows the smallest capture coefficients of the three dopants, with it being only a moderately active electron trap at the $(+1/0)$ transition, and is therefore the least detrimental to carrier lifetime. This kinetic result is consistent with the thermodynamic finding that Ca is difficult to incorporate into the CZTS lattice owing to its large ionic radius and high formation energy \cite{berman2019role}, so that Ca is expected to play only a minor role in the recombination behaviour of the alloyed material on both kinetic and thermodynamic grounds. Among the three alkaline-earth elements, Mg is thus the one whose interstitial defects most warrant attention in device optimization.

The results in \ref{sec:capture} show that  the capture rate is qualitatively consistent with the classical energy barrier height. However, the electronic coupling matrix element also plays an important role in determining capture rates, especially in low barrier cases, as exemplified by the Mg(+1/+2) transition. Despite the hole capture barrier being much lower than the electron capture barrier, the hole capture coefficient is smaller than the electron capture coefficient, due to its much weaker coupling matrix element. Nonetheless, the classical barrier can still indicate whether a capture process is fast enough for the defect to act as an active recombination center.

We observe that the barrier height qualitatively depends on both the magnitude of configuration coordinate displacement $\Delta Q$ as well as the stiffness of the potential. As previously mentioned in section \ref{sec:cc}, larger atoms tend to displace bigger, and this agrees with Ca having the largest displacements and lowest capture coefficients. However, in the Be(+1/+2) anharmonicity and a high barrier is also observed despite the lattice relaxation in Q being much smaller than the anharmonic transitions of Mg and Ca. This may be explained by the stiffer potential observed in both Be transitions, leading to higher energies on smaller Q values compared to the other dopants, which also result in higher barriers in the anharmonic case.

The differences in the bonding characteristics of the Be defect state compared to the Mg and Ca defect states might be correlated with this behaviour. As shown by fig. S11, the presence of filled Be-p orbitals indicates that a covalent $sp$ hybridization is formed with the neighboring host atoms. This interpretation is supported by Fajan's rule, which imply that the Be interstitial ion as a small cation will strongly polarize the surrounding electron clouds, resulting in covalency \cite{weller2018}. The state is not present in the Mg and Ca atoms in the other cases. implying that the electrons are donated to neighboring host atoms without forming bonds with the dopant atom itself 

More broadly, the present work extends the first-principles carrier-capture framework, previously applied to the intrinsic ``killer'' defects of CZTS \cite{killerdefect, lonepair, Kim2020upperlimit}, to extrinsic alloy-induced defects. The Sn-$5s$, Cu-$3d$, and S-$3p$ character of the alkaline-earth defect states (section ~\ref{sec:elec}) indicates that these interstitials couple to the same Sn-centred bonding network that renders the intrinsic Sn-related defects so efficient, providing a common microscopic origin for strong electron--phonon coupling in both defect classes. The influence of alloying on photovoltaic performance should therefore be understood as a competition between two processes: the reduction of intrinsic recombination-active defects through disorder suppression, and the possible introduction of recombination-active extrinsic defects. An effective alloying strategy must satisfy both criteria simultaneously, which requires screening candidate elements not only by their effect on defect formation energies but also by the capture activity of the defects they introduce.

Several approximations qualify the quantitative results while leaving the qualitative ranking intact. The PBE$+U$ functional underestimates the band gap (section~\ref{sec:elec}), which introduces uncertainty in the absolute capture coefficients through the position of the defect levels relative to the band edges; because all defects are treated identically, the relative trends are expected to be robust.  In addition, the capture coefficients reported here quantify the intrinsic capture strength of each charge-state transition; because the $+1$ state is metastable for all three dopants, the net recombination rate is further modulated by its steady-state occupation, an effect common to all three dopants that does not alter their relative ordering. Finally, the charged-defect energetics were obtained with a point-charge finite-size correction; more advanced electrostatic corrections would improve the absolute transition levels but are not expected to change the order-of-magnitude separation between the active and inactive capture channels.

%
\begin{table*}[t]
\centering
\caption{Calculated electron ($C_{\mathrm{e}}$) and hole ($C_{\mathrm{h}}$) carrier-capture coefficients of alkaline-earth interstitial defects in CZTS at $T=300$ K.}
\label{tab:capture}
\begin{tabular}{llcc}
\toprule
Defect & Charge-state transition &
$C_{\mathrm{n}}$ (cm$^{3}$ s$^{-1}$) &
$C_{\mathrm{p}}$ (cm$^{3}$ s$^{-1}$) \\
\midrule
\multirow{2}{*}{Be$_i$}
& $+1/0$ & $2.38\times10^{-9}$ & --- \\
& $+2/+1$ & $1.65\times10^{-16}$ & $1.59\times10^{-18}$ \\
\midrule
\multirow{2}{*}{Mg$_i$}
& $+1/0$ & $2.03\times10^{-12}$ & ---  \\
& $+2/+1$ & $7.14\times10^{-10}$ & $1.25\times10^{-11}$ \\
\midrule
\multirow{2}{*}{Ca$_i$}
& $+1/0$ & $1.78\times10^{-13}$ & ---  \\
& $+2/+1$ & $4.63\times10^{-15}$ & $1.84\times10^{-19}$ \\
\bottomrule
\end{tabular}
\end{table*}
%
%
\section{Conclusion}
In this work, we systematically investigated the nonradiative recombination activities of Be, Mg, and Ca extrinsic defects in kesterite Cu$_2$ZnSnS$_4$ using first-principles density functional theory combined with the nonradiative multiphonon formalism. The electronic structures, thermodynamic charge transition levels, configuration-coordinate diagrams, and temperature-dependent carrier-capture coefficients were evaluated to establish the microscopic mechanisms governing defect-assisted Shockley--Read--Hall (SRH) recombination.

Our calculations show that while all three Cu and Zn substitutionals are electronically benign, the alkaline-earth interstitials introduce deep localized defect states within the band gap and possess thermodynamic transition levels capable of participating in carrier capture. Nevertheless, the carrier-capture coefficients differ by many orders of magnitude, demonstrating that the energetic position of the defect level alone is insufficient for predicting nonradiative recombination activity. Instead, the recombination efficiency is governed by the combined effects of lattice relaxation, electron--phonon coupling, activation barriers, and vibrational overlap associated with the corresponding charge-state transitions.

Among the investigated defects, Mg$_i$ exhibits the most pronounced recombination activity. Although Be$_i$ acts as an efficient electron trap for the neutral-to-singly charged transition and Ca$_i$ remains comparatively inactive for both investigated transitions, Mg$_i$ maintains appreciable carrier-capture activity across the relevant charge-state transitions, enabling efficient sequential carrier capture and making it the most recombination-active alkaline-earth interstitial considered in this work. These results indicate that evaluating the recombination activity of a defect requires consideration of the complete sequence of charge-state transitions rather than an individual carrier-capture process.

The present results further provide a microscopic interpretation of the concentration-dependent behavior of Mg-alloyed CZTS reported experimentally. While substitutional Mg has been widely demonstrated to suppress Cu--Zn antisite disorder and improve crystal quality, the formation of Mg interstitials may introduce competing nonradiative recombination pathways that reduce the minority-carrier lifetime. Consequently, the overall influence of Mg alloying is determined by the balance between the beneficial suppression of intrinsic defects and the detrimental formation of recombination-active interstitial defects. More broadly, this work demonstrates that defect thermodynamics alone is insufficient for evaluating alloying strategies in photovoltaic absorbers. The carrier-capture coefficient constitutes a complementary kinetic descriptor that directly quantifies the impact of alloy-induced defects on nonradiative recombination. Therefore, future computational design of defect-tolerant kesterite absorbers should simultaneously consider defect formation energetics and carrier-capture kinetics in order to achieve a more comprehensive prediction of photovoltaic performance.
\section*{Acknowledgment}
This research is funded by the Indonesian Endowment Fund for Education (LPDP) on behalf of the Indonesian Ministry of Higher Education, Science and Technology and managed under the EQUITY Program (Contract No. 4298/B3/DT.03.08/2025). Several calculations in this work have been performed using the facilities of Center for Advanced Scientific Computing at Japan Advanced Institute of Science and Technology (JAIST).
\section*{Authors Contribution}
\textbf{Conceptualization}: G.S. \textbf{Investigation}: A.P.W., G.S., M.K.A; \textbf{Formal Analysis}: A.P.W, G.S., M.K.A, A.R.S.; \textbf{Writing - original draft}: A.P.W., G.S.; \textbf{Writing - review \& editing}: AP.W., G.S., M.K.A., A.R.S., K.H.,; \textbf{Supervision}: G.S., M.K.A.; \textbf{Funding Acquisition}: M.K.A., G.S., A.R.S. \\

\section*{Declaration of generative AI and AI-assisted technologies in the writing process}
 During the preparation of this work the authors used generative AI tools in order to improve the phrasing, language, and readability of the manuscript. After the use of these tools, the authors reviewed and edited the content as needed and take full responsibility for the content of the publication.

\section*{Conflicts of interest}
There are no conflicts to declare.
\printbibliography

\end{document}